\documentclass[aps,prd,twocolumn,showpacs,nofootinbib]{revtex4-1}

\usepackage{amssymb} \usepackage{amsmath} \usepackage{graphicx}
\usepackage{epsfig,latexsym}

\RequirePackage{xspace} \allowdisplaybreaks

\begin{document}

\def\bef{\begin{figure}}
\def\eef{\end{figure}}

\newcommand{\nl}{\nonumber\\}

\newcommand{\ans}{ansatz }
\newcommand{\be}[1]{\begin{equation}\label{#1}}
\newcommand{\beq}{\begin{equation}}
\newcommand{\ee}{\end{equation}}
\newcommand{\beqn}[1]{\begin{eqnarray}\label{#1}}
\newcommand{\eeqn}{\end{eqnarray}}
\newcommand{\bd}{\begin{displaymath}}
\newcommand{\ed}{\end{displaymath}}
\newcommand{\mat}[4]{\left(\begin{array}{cc}{#1}&{#2}\\{#3}&{#4}
\end{array}\right)}
\newcommand{\matr}[9]{\left(\begin{array}{ccc}{#1}&{#2}&{#3}\\
{#4}&{#5}&{#6}\\{#7}&{#8}&{#9}\end{array}\right)}
\newcommand{\matrr}[6]{\left(\begin{array}{cc}{#1}&{#2}\\
{#3}&{#4}\\{#5}&{#6}\end{array}\right)}
\newcommand{\cvb}[3]{#1^{#2}_{#3}}
\def\lsim{\raise0.3ex\hbox{$\;<$\kern-0.75em\raise-1.1ex
e\hbox{$\sim\;$}}}
\def\gsim{\raise0.3ex\hbox{$\;>$\kern-0.75em\raise-1.1ex
\hbox{$\sim\;$}}}
\def\abs#1{\left| #1\right|}
\def\simlt{\mathrel{\lower2.5pt\vbox{\lineskip=0pt\baselineskip=0pt
           \hbox{$<$}\hbox{$\sim$}}}}
\def\simgt{\mathrel{\lower2.5pt\vbox{\lineskip=0pt\baselineskip=0pt
           \hbox{$>$}\hbox{$\sim$}}}}
\def\unity{{\hbox{1\kern-.8mm l}}}
\newcommand{\eps}{\varepsilon}
\def\ep{\epsilon}
\def\ga{\gamma}
\def\Ga{\Gamma}
\def\om{\omega}
\def\omp{{\omega^\prime}}
\def\Om{\Omega}
\def\la{\lambda}
\def\La{\Lambda}
\def\al{\alpha}
\newcommand{\ov}{\overline}
\renewcommand{\to}{\rightarrow}
\renewcommand{\vec}[1]{\mathbf{#1}}
\newcommand{\vect}[1]{\mbox{\boldmath$#1$}}
\def\tm{{\widetilde{m}}}
\def\mcirc{{\stackrel{o}{m}}}
\newcommand{\Dm}{\Delta m}
\newcommand{\dm}{\varepsilon}
\newcommand{\tanb}{\tan\beta}
\newcommand{\nbar}{\tilde{n}}
\newcommand\PM[1]{\begin{pmatrix}#1\end{pmatrix}}
\newcommand{\up}{\uparrow}
\newcommand{\down}{\downarrow}
\def\omE{\omega_{\rm Ter}}
%

\newcommand{\Dsusy}{{susy \hspace{-9.4pt} \slash}\;}
\newcommand{\DCP}{{CP \hspace{-7.4pt} \slash}\;}
\newcommand{\mc}{\mathcal}
\newcommand{\gr}{\mathbf}
\renewcommand{\to}{\rightarrow}
\newcommand{\gtc}{\mathfrak}
\newcommand{\wh}{\widehat}
\newcommand{\br}{\langle}
\newcommand{\kt}{\rangle}


\def\lsim{\mathrel{\mathop  {\hbox{\lower0.5ex\hbox{$\sim$}
\kern-0.8em\lower-0.7ex\hbox{$<$}}}}}
\def\gsim{\mathrel{\mathop  {\hbox{\lower0.5ex\hbox{$\sim$}
\kern-0.8em\lower-0.7ex\hbox{$>$}}}}}

\def\nn{\\  \nonumber}
\def\de{\partial}
\def\brf{{\mathbf f}}
\def\bbf{\bar{\bf f}}
\def\bF{{\bf F}}
\def\bbF{\bar{\bf F}}
\def\bA{{\mathbf A}}
\def\bB{{\mathbf B}}
\def\bG{{\mathbf G}}
\def\bI{{\mathbf I}}
\def\bM{{\mathbf M}}
\def\bY{{\mathbf Y}}
\def\bX{{\mathbf X}}
\def\bS{{\mathbf S}}
\def\bb{{\mathbf b}}
\def\bh{{\mathbf h}}
\def\bg{{\mathbf g}}
\def\bla{{\mathbf \la}}
\def\bmu{\mathbf m }
\def\by{{\mathbf y}}
\def\bmu{\mbox{\boldmath $\mu$} }
\def\bsig{\mbox{\boldmath $\sigma$} }
\def\bunity{{\mathbf 1}}
\def\cA{{\cal A}}
\def\cB{{\cal B}}
\def\cC{{\cal C}}
\def\cD{{\cal D}}
\def\cF{{\cal F}}
\def\cG{{\cal G}}
\def\cH{{\cal H}}
\def\cI{{\cal I}}
\def\cL{{\cal L}}
\def\cN{{\cal N}}
\def\cM{{\cal M}}
\def\cO{{\cal O}}
\def\cR{{\cal R}}
\def\cS{{\cal S}}
\def\cT{{\cal T}}
\def\eV{{\rm eV}}
%

\title{More about Neutron Majorana mass from Exotic Instantons: an alternative mechanism in Low-Scale String theory }

\author{Andrea Addazi$^1$}\email{andrea.addazi@infn.lngs.it}
\affiliation{$^1$ Dipartimento di Fisica,
 Universit\`a di L'Aquila, 67010 Coppito AQ and
LNGS, Laboratori Nazionali del Gran Sasso, 67010 Assergi AQ, Italy}

\begin{abstract}

We discuss an alternative 
for Baryon-violating six quarks transitions, 
in the contest of low scale string theory. 
In particular, with
$M_{S}=10\div 10^{3}\rm TeV$,
such a transition can be mediated 
by two color-triplets,
through a quartic coupling with down-quarks,
generated by exotic instantons,
in a calculable and controllable way. 
We show how FCNCs 
 limits on color-triplet mass
are well compatible with $n-\bar{n}$ oscillation ones.
If a $n-\bar{n}$ transition were found,
this would be an indirect hint for our model.
 This would strongly motivate
 searches for direct channels in proton-proton colliders.
In fact, our model can be directly tested in a 
experimentally challenging $100\div 1000\, \rm TeV$ proton-proton collider,
searching for our desired color-triplet states
and an evidence for exotic instantons resonances,
 in addition to stringy Regge resonances,
anomalous $Z'$-bosons and gauged megaxion.
In particular, our scenario can be related to the 
$750\, \rm GeV$ diphoton hint identifying it 
with the gauged megaxion dual to the $B$-field.
On the other hand, this scenario is compatible with TeV-ish
color triplets visible at LHC and with $1\div 10\, \rm TeV$-
string scale, {\it i.e} stringy resonances 
at LHC.  

\end{abstract}

\maketitle
\section{Introduction}

The possibility that the string scale 
can be at much lower energies than 
the Planck scale $M_{P}\simeq 10^{19}\, \rm GeV$,
 is intriguing
and theoretically motivated. 
In fact, if the string scale was close to the TeV-scale, 
the hierarchy problem of the Higgs mass would be automatically solved
\cite{ADD1,AADD,ADD2,RS1,RS2}.
LHC will provide a direct test for TeV-scale string theories, 
searching for stringy Regge resonances
as well as
massive gluons, massive gravitons, mini black holes. 
On the other hand, we can argue that also if 
$M_{S}\simeq 10^{2} \div 10^{3}\, \rm TeV$, 
the hierarchy problem can be much alleviated:
$m_{H}/M_{S}\sim 10^{-(3\div 4)}$ rather than 
$10^{-14}$. In this case, direct searches at LHC 
are not possible:
stringy resonances 
would be found in
 $100\div 1000\, \rm TeV$ 
proton-proton colliders \footnote{In this scenario, also a polynomial running of cross sections with energy and the formation
of non-perturbative classical configurations could be detected in future colliders or in UHECR. 
See \cite{Addazi:2015ppa} for a recent discussion of a string-inspired effective non-local QFT unitarized by classicalization.}. 
However, effective operators 
can be generated, leading to intriguing signatures 
in low energy physics.
Recently,
we have shown, how in string-inspired models \footnote{See \cite{Addazi:2015dxa} for discussions of other different aspects about string-inspired susy QFT models.},
non-perturbative effects called {\it exotic stringy instantons}
can generate new effective operators, 
violating Baryon and Lepton numbers 
\cite{Addazi1,Addazi:2015ata,Addazi3,Addazi4,Addazi6,Addazi7,Addazi:2015goa,Addazi:2015yna}.
In particular, we have shown how
the generation of a Majorana mass term 
for the neutron from exotic instantons 
can be possible, without proton destabilization. This leads to the possibility
to test indirectly this class of models, in the next generation 
of experiments on neutron-antineutron oscillations. 
The actual best limits on $n-\bar{n}$ transition is only $\tau_{n\bar{n}}\simeq 3\, \rm yr$
\cite{NNbar1,NNbar2}, and the next generation of experiments 
will enhance this one by two orders of magnitude \cite{NNbar2}. 
In \cite{Addazi1,Addazi:2015ata,Addazi3,Addazi4,Addazi6,Addazi7,Addazi:2015goa,Addazi:2015yna}, we have shown how proton is not destabilized;
but  deviations in neutral mesons oscillations and other FCNC
are generically predicted, roughly at the same scale 
as for $n-\bar{n}$ oscillations. 
As a sequel of our paper, we propose
a new alternative mechanism
 for the generation of a Neutron Majorana mass from exotic instantons, as a variant of the ones discussed in \cite{Addazi1,Addazi:2015ata,Addazi3,Addazi4}\footnote{
In our papers, we have considered a class of exotic instantons wrapping different 3-cycles with respect to 
ordinary D-branes. On the other hand, also different classes of exotic instantons, studied in \cite{Parsa1,Parsa2,Parsa3}, may be relevant for phenomenology. }.
 In
 low scale string theory, $M_{S}\simeq 10\div 10^{3}\, \rm TeV$,
this mechanism can be tested indirectly in $n-\bar{n}$ experiments 
and FCNC processes.  
 In particular, we propose a
 general class of 
intersecting D-brane models, 
in which the SM is embedded,
and
extra color-triplet superfields $C,C^{c}$ 
naturally emerge for construction. 
Exotic instantons generate an extra quartic
superpotential for $\mathcal{W}_{np}=CCD^{c}D^{c}/\mathcal{M}_{0}$. As a consequence, 
 a neutron-antineutron transition is induced by non-perturbative 
 effects in a theory of quantum gravity!
 \footnote{For other ideas about neutron-antineutron 
 oscillations in Large Extra Dimensions, 
 see \cite{Nussinov}.}

\section{Neutron Majorana mass from exotic instantons}

Let us consider, at effective level, a (N)MSSM
plus two extra superfields $C$ and $C^{c}$ \footnote{Even 
if our model is $\mathcal{N}=1$ supersymmetric, supersymmetry is not necessarly 
broken at TeV-scale: it can be broken close to the string-scale.}.
$C$ is a $(\bar{3},1)_{Y=-2/3}$ with respect to 
$SU(3)_{c}\times SU(2)_{L}\times U(1)_{Y}$,
with a Baryon number $B(C)=-2/3=2B(D^{c})$. 
We can introduce at perturbative level the
following R-parity preserving superpotential terms  \footnote{
As commented in \cite{Addazi3,Addazi4}, 
extra mass parameters, such as soft susy breaking 
ones, can be generated 
by RR-RR or NS-NS three-forms fluxes in the bulk.
In our case, mass term in the superpotential like 
$m_{C}CC^{c}$ can be generated by fluxes.
For recent discussions of 
mass deformed quivers and dimers see also \cite{Bianchi:2014qma}.} 
\be{super1}
\mathcal{W}_{p}=\mathcal{W}_{(N)MSSM}+y_{1}C_{ij} Q^{i}Q^{j}
\ee
At non perturbative level we can generate through exotic instantons 
the following extra term for $C^{c}$:
\be{extra}
\mathcal{W}_{np}=\frac{1}{\mathcal{M}_{0}}\epsilon^{ijk} \epsilon^{lmn} D^{c}_{i}D^{c}_{l}C_{jk}C_{mn}
\ee
Superpotential term (\ref{extra}) violates the baryon number as $\Delta B=2$.
The superpotentials (\ref{super1})-(\ref{extra}) generate two possible 
relevant diagrams for neutron-antineutron oscillations,
shown in Fig-1-(a)-(b). 
In the first one, relevant operators in the lagrangian are
$y_{1}\phi_{C}q_{L}q_{L}$, 
$m_{\phi_{C}}^{2}\phi_{C}\phi_{C}^{\dagger}$,
and 
$d^{c}d^{c}\phi_{C}\phi_{C}/\mathcal{M}_{0}$
and their hermitian conjugates. 
Integrating out $\phi_{C}$,
we obtain the effective operator
$\mathcal{O}_{n\bar{n}}=Tr[y_{1}y_{1}^{\dagger}](qqd^{c})^{2}/\Lambda_{n\bar{n}}^{5}$
where $\Lambda_{n\bar{n}}^{5}=m_{\phi_{C}}^{4}\mathcal{M}_{0}$. 
The diagram in Fig.1-(b) generates an effective operator $\mathcal{O}_{n\bar{n}}$
with a NP scale $\Lambda_{n\bar{n}}^{5}=\mathcal{M}_{0}m_{C}^{2}m_{\tilde{g}}^{2}$, where $m_{\tilde{g}}^{2}$ is the gaugino mass (gluino, zino or photino).
Which one of the two diagrams dominates, depends 
on the particular region of the parameters considered.
In Fig.1-(c) we show mixed disk amplitudes generating the 
effective superpotential term (\ref{extra}). We will return 
later on to the precise calculation of the relevant string amplitudes. 
\begin{figure}[t]
\centerline{ \includegraphics [height=8cm,width=1.0 \columnwidth]{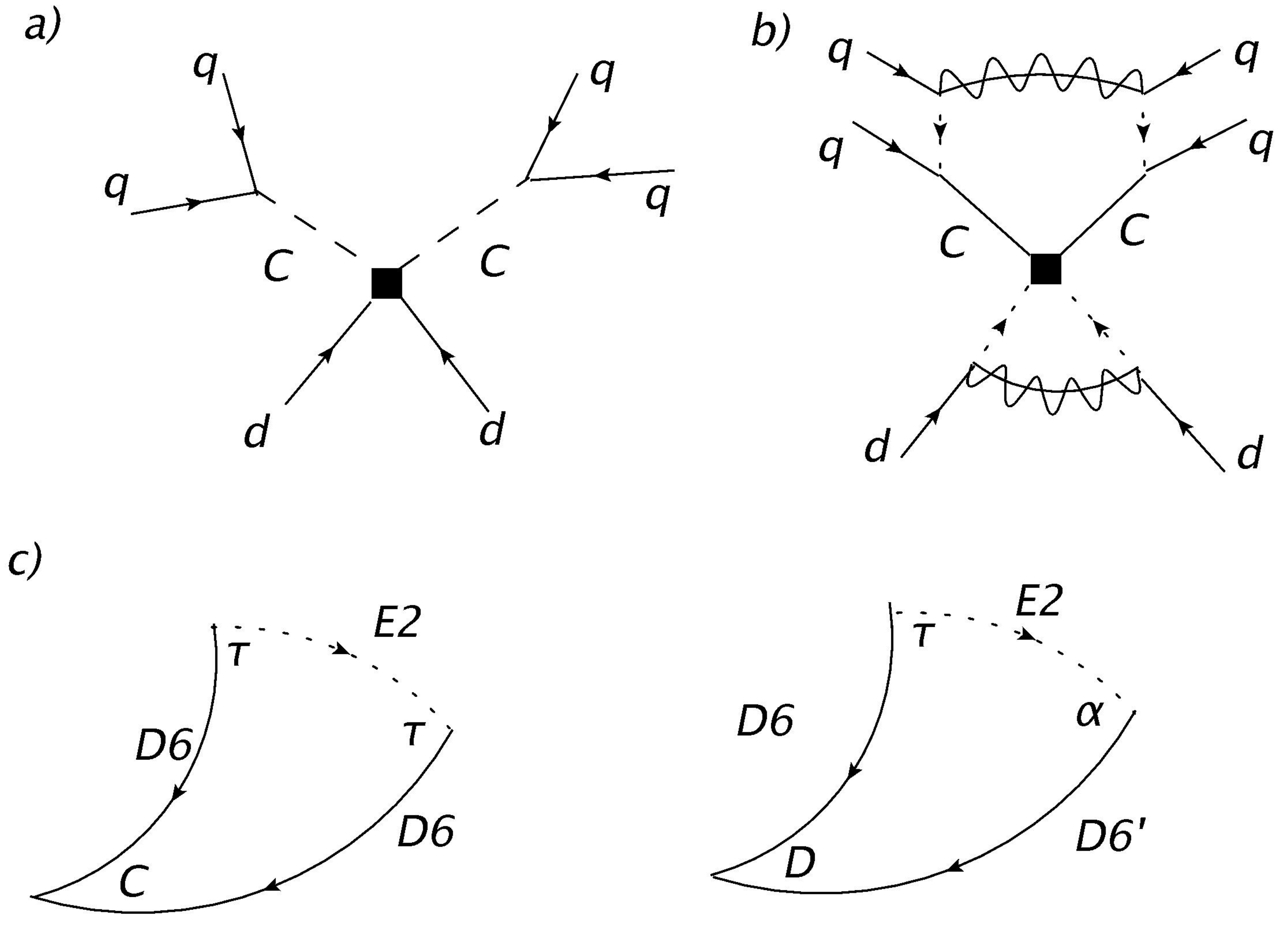}}
\vspace*{-1ex}
\caption{a) Diagram inducing neutron-antineutron transitions through
color-triplet scalars $\phi_{C}$ and an effective interaction induced by 
exotic instantons (black box). b) Supersymmetric diagram inducing neutron-antineutron transitions: two fermionic superpartners $\psi_{C}$ 
mediate the process, four quarks are converted into 
four squarks 
 through two gaugini.
(Notation: $d,D\equiv d^{c},D^{c}$ and $q\equiv q_{L}$ so that $\frac{1}{2}\epsilon_{\alpha \beta}q^{\alpha}q^{\beta}=u_{L}d_{L}$).
c) Mixed disk amplitudes inducing the relevant effective interaction between $D^{c}$ and $C^{c}$.}
\label{plot}   
\end{figure}
Now let us discuss the case in which Fig.1-(a) is dominant with respect 
Fig.1-(b), {\it i.e} supersymmetry is not related to the hierarchy problem of the Higgs mass in this case. 
Under this assumption, the relevant contributions to FCNCs
are shown in Fig.2.
In particular, 
contributions to neutral mesons oscillations as 
 $K^{0}-\bar{K}^{0}$ are generated 
by box-diagrams. Among all experimental constraints, the strongest one
comes from $K^{0}-\bar{K}^{0}$: 
this process
is suppressed as $\Lambda_{K^{0}\bar{K}^{0}}^{2}\simeq (10^{2}\, \rm TeV)^{2}$ \cite{PDG}.
Assuming $y_{1}\simeq 10^{-2}\div 1$, we can estimate 
a bound for $m_{\phi_{C}}$ as $m_{\phi_{C}}\simeq 1\div 1000\, \rm TeV$. 
On the other hand, a $b\rightarrow s\gamma$ transition is also generated: the experimental suppression approximatively puts the same limits on $m_{\phi_{C}}$. 
These bounds are higher than the direct ones coming from LHC (roughly 
near $1\, \rm TeV$ for $y_{1}\sim 1$, as discussed in \cite{Addazi:2015ata}). 
\begin{figure}[t]
\centerline{ \includegraphics [height=3cm,width=1.1 \columnwidth]{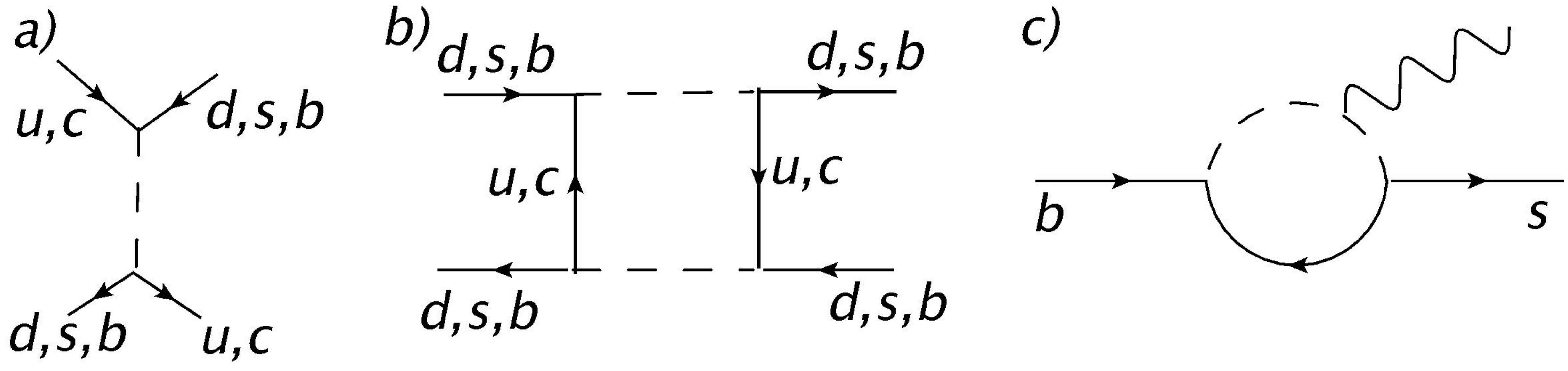}}
\vspace*{-1ex}
\caption{a) FCNCs tree-level diagrams mediated by $\phi_{C}$ (dashed lines). 
b) Diagrams of neutral-meson oscillations, mediated by two $\phi_{C}$. 
c) Diagrams for $b\rightarrow s\gamma$ transition generated by a loop of up-type quarks and a $\phi_{C}$.}
\label{plot}   
\end{figure}
We can combine these bounds with the ones from $n-\bar{n}$ oscillation,
$\Lambda_{n\bar{n}}>300\, \rm TeV$,
and the LHC one on the string scale $M_{S}>1\div 10\, \rm TeV$ \cite{Exotic}.
The next generation of experiments in $n-\bar{n}$ oscillations promise to 
test the $1000\, \rm TeV$ scale. This can correspond to
(assuming $y_{1}\simeq 1$) $\mathcal{M}_{0}\simeq m_{\phi_{C}}\simeq 10^{3}\, \rm TeV$. In this case, FCNC bounds are satisfied. Such a situation can be easily obtained trough exotic instantons. 
In case in which the diagram in Fig.1-(b)
is not sub-dominant, 
analogous bounds on FCNCs can be obtained, with one more free parameter with than in the one in Fig.1-(a):
the gaugino mass. \footnote{
As proposed recently, an auto-concealment of susy can be possible 
in extra dimensions, with effective Planck scale or stringy scale $1-100\, \rm TeV$  \cite{Dimopoulos:2014psa}. 
This scenario generically implies
large missing energy channels at LHC. 
Our proposal can be considered in this contest, 
in which the supersymmetric diagram is expected to be dominant 
or at least relevant.}.

Let us discuss the mixed disk amplitudes pictures in Fig.1-(c),
where string theory enters in our model. 
$C$ and $Q$ are excitations of open strings stretched between 
$D6_{3}-D6_{3}$ branes and $D6_{3}-D6'_{2}$-branes respectively. 
In mixed disk amplitudes, the relevant $E2$-brane instanton intersects
two times the $D6$ and $D6'$ branes. Fermionic modulini 
$\tau^{i},\alpha$
are excitations of open string attached to $D6-E2$
and $D6'-E2$ respectively. 
As a consequence the following effective 
interactions are generated:
\be{effectiveMod}
\mathcal{L}_{eff}\sim D^{c}_{i}\tau^{i}\alpha+C_{ij}\tau^{i}\tau^{j}
\ee
We consider a number of intersections of $E2$-brane and $D6_{3}-D6'_{2}$ is equal to 
\be{intersections}
I_{E2-D6_{3}}=-I_{E2-D6_{2}'}=2
\ee
Integrating over the modulini space (as usually done for istantonic solutions), we obtain 
\be{integration}
\mathcal{W}_{np}=e^{-S_{E2}}\int d^{6}\tau d^{2}\alpha e^{\mathcal{L}_{eff}}=\frac{e^{-S_{E2}}}{M_{S}}\epsilon^{ijk} \epsilon^{lmn} D^{c}_{i}D^{c}_{l}C_{jk}C_{mn}
\ee
So, $\mathcal{M}_{0}=M_{S}e^{+S_{E2}}$, where
$S_{E2}$ is the effective action of the $E2$-brane,
depending on K\"ahler moduli associated to
 3-cycles of $E2$-brane to the Calabi-Yau $CY_{3}$.
Intuitively, if 3-cycles size is small, $e^{S_{E2}}\sim 1$, 
 on the contrary $e^{S_{E2}}$ can be much higher than one.
 For example, a situation in which $\mathcal{M}_{0}\simeq 10^{3}\, \rm TeV$
 can correspond to $M_{S}\simeq 10^{3}\, \rm TeV$ ($e^{S_{E2}}\sim 1$)
 as well as to $M_{S} \simeq 10\, \rm TeV$ ($e^{S_{E2}}\sim 10^{2})$.
We note that exotic instantons have dynamically violated
R-parity and the Baryon number, without generating 
proton or neutralino decays operators. 
 
\begin{figure}[t]
\centerline{ \includegraphics [height=5cm,width=0.7 \columnwidth]{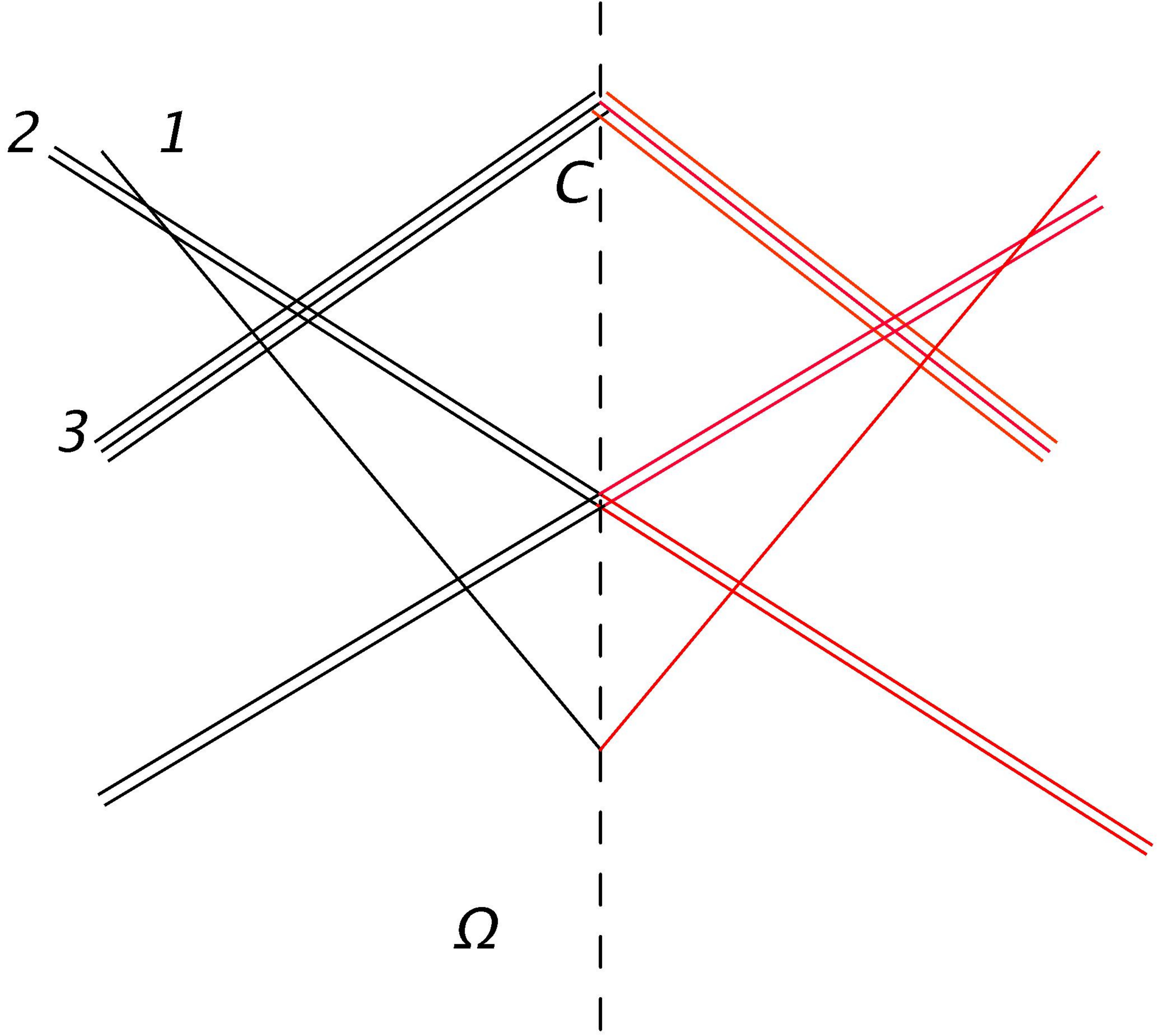}}
\vspace*{-1ex}
\caption{We show a simplified scheme of a possible class of D-brane systems,
generating the fields content of our model.
In particular, system shown in figure reproduces $G_{NMSSM}=U(3)\times U(2) \times U(1)$.
Clearly, $G_{NMSSM}$ can be extended with one 
(or more extra) $U(1)$.
An antisymmetric $\Omega_{-}$-plane is introduced. 
$C,C^{c}$ are obtained as excitations of open strings attached to $U(3)_{c}$
and its mirror reflection.
Notation: $3\equiv U(3)_{c}$, $2\equiv U(2)$, $1\equiv U(1)$,
in black ordinary D-branes, in red "images" of physical D-branes.  }
\label{plot}   
\end{figure}

The effective model proposed
can be embedded in several different
classes of (NM)SM-like intersecting D-brane systems.
See \cite{DB1,DB2,DB3,DB4,DB5,DB6,DB7,DB8,DB9,DB10,DB11,DB12,DB13,DB14,DB15,DB16,DB17}
for papers and reviews in these subjects, with particular emphasis 
to possible implications for LHC. 
A minimal choice can be to consider
D-branes systems for $U(3)\times U(2) \times U(1)$.
In our construction, we introduce an antisymmetric 
Mirror $\Omega^{-}$-plane,
as shown in Fig.3. 
This allows to construct $C,C^{c}$ from open strings attached between $U(3)_{c}$ and its mirror twin.
In particular, 
one stack of three D6-branes produces the $U(3)_{c}$ gauge group,
including $SU(3)_{c}$ and an extra $U(1)$;
one stack of two D6-branes
for $U(2)_{L}$, containing $SU(2)_{L}$ and an extra $U(1)$,
one stack of a single D6-brane for a $U(1)$ gauge group;
an antisymmetric $\Omega$-plane, identifies the D-brane stacks 
with their images. Let us remind the, $\Omega$-planes explicitly breaks $\mathcal{N}=4$ theory
down to $\mathcal{N}=1$ susy theory, and they
are usually introduced for tadpole cancellations
 \cite{Sagnotti1,Sagnotti2,Sagnotti3,Sagnotti4,Sagnotti5,Sagnotti6,Sagnotti7,Sagnotti8, sessantatre, sessantaquattro, MBJFM, Angelantonj:1996uy, Angelantonj:1996mw}
\footnote{Another important element for intersecting D-brane models can be flavor branes \cite{18,19,20,BIMP},
but we do not discuss possible explicit model with these ones.}.
The presence of orientifold planes $\Omega$
in intersecting D-brane system seems a key element  
for realistic models of particle physics. 
Let us note that $U(1)_{3}\subset U(3)_{c}$ and $U(1)_{2}\subset U(2)_{L}$
are anomalous in gauge theories. On the other hand, 
these $U(1)$s are not problematic in string theories.
In fact, a generalized Green-Schwarz mechanism can cancel 
 anomalies, through generalized Chern-Simon 
(GCS) terms. 
The new vector bosons $Z',Z''$ associated to $U(1)_{2,3}$
 get masses via St\"uckelberg mechanisms.
 See  \cite{Stuck1,Stuck2,Stuck3,Stuck4,Stuck5,Stuck6,Stuck7,Stuck8,Stuck9,Stuck10,Stuck11,Stuck12} for discussions about these aspects in different contests \footnote{Let us mention that another implementation of
 the St\"uckelberg mechanism is in the realization of a Lorentz Violating
 Massive gravity \cite{LIV1,LIV2,LIV3}. Recently, geodetic instabilities of St\"uckelberg
 Lorentz Violating
 Massive gravity were discussed in \cite{Addazi:2014mga}. In subregions 
 of parameters' space of these models, naked singularities are allowed.
 This could be connected to the existence of new items called {\it frizzyballs}
 in these theories   \cite{Addazi:2015gna,Addazi:2015hpa,Addazi:2015cho,Addazi:2016cad}. 
  }
 \footnote{Another intriguing application of exotic instantons 
 regards the generation of RH neutrini masses and $\mu$-terms
 \cite{Blu1,Blu2,Ibanez1,Ibanez2,Abe:2015uma}. 
  This idea is compatible 
 with our one.}.
 Finally, hypercharge $U(1)_{Y}$ is a non-anomalous linear combination 
 of $U(1)_{3}$, $U(1)_{2}$, $U(1)_{1}$ of Fig.3. 
Our idea is so generic that 
 can be implemented in several different D-brane models.
The precise hypercharge combination depends on the particular 
D-brane construction considered.
However, let us note that 
a complete classification of SM-like 
D-branes' models
are 
 in \cite{Jimi1,Jimi2,Jimi3}. In these models, 
 the presence of extra exotic matter $C,C^{c}$ often seems 
 necessary for a consistent cancellation of all tadpoles and anomalies!
 So, one can just consider these models in presence of an opportune $E2$-brane 
 like the one suggested here. Clearly, this un-balances in/out-oriented strings for each stack. 
However, one can introduce a flavor brane so that the number
of in/out oriented strings remains zero for each stack. 
 The construction of a precise quiver theory is not the purpose
 of this paper, even if we think that just the minimal extension
 described just above is sufficient. 
 On the other hand, our "$\Omega$-trick" can be considered 
 also in D-brane constructions for models like 
 3-3-1 as the one considered in \cite{Valle1,Valle2,Valle3}.
 Usually 3-3-1 models as the one considered 
 in \cite{Valle1,Valle2,Valle3} are not easily embedded in 
 GUT-inspired models, for its peculiar fields content,
 while in intersecting D-brane ones
 there are less difficulties to get such a model, as shown in Ref. \cite{Addazi:2016xuh}\footnote{I would like to thank 
 Jos\'e Valle for useful comments on these subjects.}
 \footnote{Let us also mention that in contest of intersecting D-branes models, 
 an interpretation of dark matter and dark energy as hidden sector is particularly
 motivated. Recently scenari in which dark matter and dark energy are unified by a hidden strong sector
 were suggested in Ref. \cite{Addazi:2016sot,Addazi:2016nok}. In these references, we also commented about possible 
 connections with intersecting D-branes models.  }. 
Finally, an  $E2$-brane is introduced, intersecting 
 stacks in Fig.3 as indicated in Fig.1-(c),
 and
  generating our {\it desiderata} superpotential (\ref{extra}).

\section{Implications for high energy proton-proton colliders}

In this section, we will comment on further implications of 
our model for high energy colliders. 

In particular, the generation of the effective six-quark operator 
generated in our model has
flavor matrix structure. 
In other words, integrating out $C,C^{c}$ fields for $E<<M_{C}$, 
our model not only predicts $udd\rightarrow \bar{u}\bar{d}\bar{d}$ transitions, 
but also $uds \rightarrow \bar{u}\bar{d}\bar{s},...,bus\rightarrow \bar{b}\bar{u}\bar{s}$ and so on. 
This implies that for $E_{CM}\simeq M_{C}$, color sextets can decay into two (anti)quarks as 
$$C^{c}\rightarrow udd,\,\,\,C\rightarrow \bar{u}\bar{d}\bar{d},\,\,...\,\,C^{c}\rightarrow bus,\,\,\,C\rightarrow \bar{b}\bar{u}\bar{s},\,\,...\,\,$$
In general each of these processes are controlled by different couplings, 
geometrically understood by the mixed disk amplitude structure
related to the exotic instanton solution. 
So that for $M_{C}\simeq 10\div 100\, \rm TeV$, colored triplets decays can be discovered or limited at
LHC or $\sqrt{s}=100$-TeV colliders beyond LHC. 
However, direct research limits have to be compared with indirect measures of FCNC processes
as shown before. For instance 
$y_{1,Cud}Cu_{L}d_{L}$,
$y_{1,Ccd}Cc_{L}d_{L}$,
 $y_{1,Cus}Cu_{L}s_{L}$ and 
$y_{1,Ccs}Cc_{L}s_{L}$ operators are constrained 
approximately as
$y_{1,Cud}y_{1,Cus}\simeq y_{1,Ccd}y_{1,Ccs}\simeq (100\, \rm TeV/m_{C})^{2}$. 
So that, for $m_{C}\simeq 1\div 10\, \rm TeV$, 
limits from LHC on $C\rightarrow \bar{u},\bar{d},\bar{u}\bar{s},\bar{c}\bar{d},\bar{c}\bar{s}$
are less stringent than limits from FCNCs. 
However, other diagrams involving other generation of quarks 
are less constrained and processes like 
$C\rightarrow \bar{t} \bar{b},\bar{c}\bar{b}$ can be detected by LHC
with coupling of the order one. 
In fact, they can generate $K_{0}-\bar{K}_{0}$ only by extra CKM electroweak loop reductions, i.e., 
depending on the generation involved, an extra suppression of $10^{-2}\div 10^{-6}$ for the transition amplitude. 
As regards the exotic instanton, for an effective scale $\Lambda\simeq M_{S}e^{+S_{E2}}$.
A direct test of our model is the detection of the exotic instantons in collisions.
As we discussed for 
 $E<<\mathcal{M}_{0} \simeq M_{S}e^{+S_{E2}}$, the $E2$-brane is rigidly intersecting 
 the ordinary $D6$-branes, generating an effective contact interaction. 
 This means that the cross section of the process 
 $qq\rightarrow \bar{q}\bar{q}\bar{q}\bar{q}$ 
 is 
 \be{sigmaqq}
 \sigma(qq\rightarrow \bar{q}\bar{q}\bar{q}\bar{q})=\left(Tr[y_{1}y_{1}^{\dagger}]\right)^{2}\frac{s^{3}}{\mathcal{M}_{0}^{2} m_{C}^{8}}
 \ee
 for $\sqrt{s}<\left(\mathcal{M}_{0} m_{C}^{4}\right)^{1/5}$. 
 For $\sqrt{s}>> \left(\mathcal{M}_{0} m_{C}^{4}\right)^{1/5}$, 
 the cross-section (\ref{sigmaqq}) is not still valid.
 For instance, it would violate unitarity. 
At that scale, the $E2$-brane cannot more be considered a rigid E-brane and 
its oscillations are described by fermionic modulini $\tau$ and $\alpha$. 
At the same scale, the fermionic modulini reggeaize: from the first massless modulini 
massive Regge states are inevitably excited at this scale, 
with a Regge slope $\alpha=-1/2+\alpha' s$. 
In the region of high energy scattering 
where all Mandelstam variables are higher than the string scale $|s_{ij}|>>M_{S}^{2}$ 
and fixed ratios $s_{ij}/s={\rm const}$, 
the amplitude will have the universal stringy exponential suppression \cite{Gross:1987ar}, 
\be{A6}
\mathcal{A}_{6}\sim e^{-\sum_{ij}\alpha' s_{ij} {\rm log} \alpha' s_{ij}}
\ee
Let us note that also at the non-perturbative scale the E2-brane 
has to conserve its number of intersections which are topological invariants 
of the mixed disk amplitudes. 
The production of an exotic instanton in collision 
is completely different by the production of an elementary particle:
while a elementary resonance has a Breit-Wigner amplitude peak, 
the production of an exotic instanton is expected to be correspond to a highly asymmetric resonance peak: 
before $\Lambda$ scale the cross section is polynomially increasing while 
after the cutoff scale $\Lambda$ the cross section is exponentially softened.

Now let us comment on other implications for colliders not 
directly related to our mechanism for a neutron-antineutron transition 
but inevitably predicted in the class of models considered.
 In fact, the presence of a supersymmetric scale and 
 a low string scale inevitably leads to the presence 
 of supersymmetric particles and higher spins Regge states. 
 On the other let us note that recent results of LHC 
disfavor  scenari like 
 $M_{S}\simeq 1\div 10\, \rm TeV$
and $M_{SUSY}\simeq 1\div 10\, \rm TeV$.
In the contest of our model, the favored region of parameters 
is $M_{S}\simeq M_{SUSY}\simeq 100\, \rm TeV$, 
more naturally compatible with FCNCs and Neutron-Antineutron limits. 
In this scenario, colored supersymmetric partners like 
gluini, squarks and susy fermionic sextets $\psi_{C}$
are expected to be detectable in $100\, TeV$ proton-proton colliders. 
On the other hand, massive Regge states like massive gluons, massive 
gravitons or massive $B_{\mu\nu}$-fields are generically predicted
in low scale string theory models.  
As mentioned in the previous section, the presence of new massive 
abelian gauge bosons $Z'$ interacting through generalized Chern-Simon terms 
is an avoidable prediction of our class of models. 
For instance they can decay at tree-level as $Z'\rightarrow Z\gamma$, producing
a CP-violating polarized photon. 
Unitarity of $Z'$ amplitude imposes the constrain $m_{Z'}\simeq M_{S}$. 
In fact, one-loop correction using vertices $Z'Z\gamma$ leads to quadraticaly UV divergent 
unitarized at the string scale $M_{S}$ \footnote{We thank Massimo Bianchi for private communications 
on these results.}. 
So that, $Z'$ can be detected at LHC if and only if $M_{S}\simeq 10\, \rm TeV$. 
For a scenario $M_{S}\simeq 100\, \rm TeV$, $Z'$ bosons can be tested in 
new proton-proton colliders beyond LHC. 

Finally, a related implications of the class of model under consideration is the 
presence of closed strings rank 2 antisymmetric field $B_{\mu\nu}$ dual to 
a massive axion as ${\bf B}={\bf d} {\bf b}$. 
In low string scale scenario, the $750\, \rm GeV$ hint 
measured by ATLAS and CMS can be interpreted as
such a gauged axion \cite{Ibanez:2015uok,Anchordoqui:2016rve,Li:2016tqf,Anastasopoulos:2016cmg,Anchordoqui:2016rve}

\section{Conclusions and discussions}

In this paper, we have seen a model 
for the generation of a Majorana mass from Exotic stringy instantons,
as a variant of models proposed in \cite{Addazi1,Addazi2,Addazi3,Addazi4}.
We have discussed a scenario in which new color-triplet states
interact with ordinary quarks, mediating a neutron-antineutron transition.
The key of the mechanism is the generation of 
 a non-perturbative quartic interaction 
 among color-triplets and quarks
 from Exotic Instantons. 
Proton is not destabilized in this model
as well as neutralino or other possible LSP.
As a consequence, neutron-antineutron transition can be fast 
as $\tau_{n\bar{n}}\simeq 300\, \rm yr$ ($1000\, \rm TeV)$, 
in a low scale string theory scenario
$M_{S}=10\div 10^{3}\, \rm TeV$. 
This scenario can be well compatible 
with FCNC limits,
in particular with $K_{0}-\bar{K}_{0}$ and $b\rightarrow s\gamma$.
Finally we discussed how this scenario motivates 
direct researches of colored triplets and 
exotic instantons in high energy proton-proton colliders. 
In the Early Universe, 
new decays $\phi_{C}\rightarrow \bar{q}\bar{q},qq$, also counting one-loop contributions,
could generate a Baryon asymmetry (BAU). 
However,
such a processes could be not enough efficient because of
  washing-out collisions, generated by exotic instantons, like
$d^{c}d^{c} \rightarrow \phi_{C}\phi_{C}$, and by sphalerons successively.
On the other hand, string theory suggests that all couplings are
dynamical degrees of freedom, stabilized by 
fluxes and instantons. 
 The scale generated by exotic instantons 
depends on geometric moduli,
associated to the shape of E2-brane cycles wrapping the Calabi-Yau.
As a consequence, E2-branes K\"ahler moduli 
can evolve as dynamical degrees of freedom,
so that 
$e^{-S_{E2}}(modulini)[t_{Early-Universe}]<<e^{-S_{E2}}(modulini)[t_{Present-Epoch}]$.
This corresponds to a dynamical enlargement of 3-cycles
radii, wrapped in the Calabi-Yau,
during the cosmological evolution.
For example, geometric moduli  
can be stabilized to a "solitonic solution" 
of the Cosmological time $t$, $\mathcal{F}(t)=e^{-S_{E2}}(t)$,
connecting two asymptotic branches $\mathcal{F}_{1}=\mathcal{F}(t_{Early-Universe})$
and $\mathcal{F}_{2}=\mathcal{F}(t_{Present-Epoch})>>\mathcal{F}_{1}$.
This is plausible, also considering that
usually the dependence on moduli 
is exponential-like, while a soliton solution 
is usually a combination of hyperbolic functions,
{\it i.e} a combination of exponentials. 
In this way, exotic instantons
effects are strongly suppressed 
during the Early Universe, 
not washing-out 
baryon asymmetries, generated after inflation.
On the other hand, neutron-antineutron oscillations
remain reachable for the next generation of experiments. 
Such a hypothesis deserves future 
theoretical and numerical investigations in string phenomenology
and baryogenesis calculations.

We conclude that exotic instantons continue to surprise with their intriguing implications 
in phenomenology. This strongly motivates  
researches on neutron physics and high energy colliders beyond LHC. 
This also highly motivates theoretical researches of consistent quivers and Calabi-Yau singularities 
for the class of D-brane models proposed in our letter, 
as already done in similar cases \cite{DMSSM12,DMSSM13,DMSSM14,DMSSM15}.

\vspace{1cm} 

{ \bf Acknowledgments} 
\vspace{3mm}

I would like to thank the anonymous referee, Massimo Bianchi, Gia Dvali, Cesar Gomez, Jos\'e Valle and Gabriele Veneziano for 
useful discussions on these aspects. 
I also would like to thank Antonino Marcian\'o and Fudan University for hospitality during the completion of this paper. 
My work was supported in part by the MIUR research grant Theoretical Astroparticle Physics PRIN 2012CP-PYP7 and by SdC Progetto speciale Multiasse La Societ\'a della Conoscenza in Abruzzo PO FSE Abruzzo 2007-2013.

\end{document}